\documentclass[12pt]{elsart}

\input{epsf.tex}

\def\la{\mathrel{\mathchoice {\vcenter{\offinterlineskip\halign{\hfil
$\displaystyle##$\hfil\cr<\cr\sim\cr}}}
{\vcenter{\offinterlineskip\halign{\hfil$\textstyle##$\hfil\cr
<\cr\sim\cr}}}
{\vcenter{\offinterlineskip\halign{\hfil$\scriptstyle##$\hfil\cr
<\cr\sim\cr}}}
{\vcenter{\offinterlineskip\halign{\hfil$\scriptscriptstyle##$\hfil\cr
<\cr\sim\cr}}}}}
\def\ga{\mathrel{\mathchoice {\vcenter{\offinterlineskip\halign{\hfil
$\displaystyle##$\hfil\cr>\cr\sim\cr}}}
{\vcenter{\offinterlineskip\halign{\hfil$\textstyle##$\hfil\cr
>\cr\sim\cr}}}
{\vcenter{\offinterlineskip\halign{\hfil$\scriptstyle##$\hfil\cr
>\cr\sim\cr}}}
{\vcenter{\offinterlineskip\halign{\hfil$\scriptscriptstyle##$\hfil\cr
>\cr\sim\cr}}}}}

\begin{document}
\begin{frontmatter}
\title{The power-law behaviours of angular spectra 
of polarized Galactic synchrotron
}
\author [astrofi]{M. Bruscoli\thanksref{emailbru}},
\author [milanobi]{M. Tucci\thanksref{emailtuc}},
\author [caismi]{V. Natale\thanksref{emailnat}},
\author [boh]{E. Carretti\thanksref{emailcar}},
\author [unifi]{R. Fabbri\thanksref{emailfab}},
\author [boh]{C. Sbarra\thanksref{emailsba}},
\author [boh]{S. Cortiglioni\thanksref{emailpi}}
\address [astrofi] {Dipartimento di Astronomia, Universit\`a di Firenze,
Largo Fermi 2, I-50125 Firenze, Italy}              
\address [milanobi]{Dipartimento di Fisica, Universit\`a
di Milano-Bicocca, Piazza della Scienza 3, I-20126 Milano, Italy} 
\address [caismi] {I.R.A./C.N.R., Sezione di Firenze,
Largo Fermi 5, 
I-50125 Firenze, Italy}
\address [boh] {I.A.S.F./C.N.R., Sezione di Bologna,
Via P. Gobetti 101,
I-40129 Bologna, Italy}     
\address [unifi]{Dipartimento di Fisica, Universit\`a di Firenze, 
Via Sansone 1, I-50019 Sesto Fiorentino, Italy}     

\thanks[emailbru]{E-mail: bruscoli@arcetri.astro.it}                                 
\thanks[emailtuc]{E-mail: tucci@ifca.unican.es}                              
\thanks[emailnat]{E-mail: natale@arcetri.astro.it}                                   
\thanks[emailcar]{E-mail: carretti@tesre.bo.cnr.it}                                   
\thanks[emailfab]{E-mail: fabbrir@unifi.it}                                
\thanks[emailsba]{E-mail: sbarra@tesre.bo.cnr.it}  
\thanks[emailpi]{E-mail: cortiglioni@tesre.bo.cnr.it}

\begin{abstract}
We study the angular power spectra of polarized Galactic 
synchrotron in the
range $10\la l\leq 800$, at several frequencies
between 0.4 and 2.7 GHz and at
several Galactic latitudes up to near the North Galactic Pole. 
Electric- and magnetic-parity polarization spectra
are found to have slopes around 
$\alpha _{E,B} = 1.4 - 1.5$ in the Parkes and
Effelsberg Galactic-Plane surveys, but strong local fluctuations
of $\alpha _{E,B}$ are found at $ | b | \simeq 10 ^\circ $ from
the 1.4 GHz Effelsberg survey. 
The $C_{PIl}$ spectrum,
which is insensitive to the polarization direction, is somewhat
steeper, being $\alpha _{PI} = 1.6 - 1.8$ for the same surveys. 
The low-resolution multifrequency survey of Brouw and Spoelstra
(1976) shows some flattening of the spectra below 1 GHz,
more intense for $C_{E,Bl}$ than for $C_{PIl}$. In no case we find
evidence for really steep spectra.  The extrapolation to the
cosmological window shows that at 90 GHz the detection of E-mode
harmonics in the cosmic background radiation should not
be disturbed by synchrotron, even around $ l \simeq 10$ for 
a reionization optical depth $\tau  _{\mathrm {ri}}\ga 0.05$.
\end{abstract}

\end{frontmatter}
                                              

\section{Introduction}

Angular power spectra (henceforth, APS) of the Galactic synchrotron
polarized radiation are raising up an increasing attention in these years.
The first study, based on the Parkes 2.4 GHz survey (Duncan et al. 1997,
hereafter D97), was due to Tucci et al. (2000), and other papers appeared
shortly after (Baccigalupi et al., 2001a; Giardino et al. 2001). The main
motivation for these works was the need for an 
angular-scale--dependent separation of
the cosmic microwave background (CMB) signal from the polarized radio
foreground; in fact the first work on the APS of a polarized Galactic
foreground (Prunet et al. 1998) modelled thermal dust emission in view of
the scenario to be met by Planck HFI at 143-217 GHz. CMB polarization is
essential in order to remove degeneracies between important cosmological
parameters (Zaldarriaga et al. 1997), 
but the cosmological window (the region in the
frequency--angular-scale plane where the cosmological signal is stronger
than foregrounds) is narrower for polarization than for anisotropy. A
careful study of Galactic contamination versus angular scale is thereby
necessary. The study of synchrotron APS may also be important because of its
bearing on the knowledge of Galactic structure, and in particular of the
transverse magnetic field in emission regions and the longitudinal field in
compact foreground screens (Tucci et al. 2001). This point is open to future
studies.

The purpose of this work is to check the generality of the APS behaviour
found by Tucci et al. (2000): Electric and
magnetic parity APS are governed (with reasonable approximation for $100\leq
l\leq 800$) by power laws $C_{E,Bl}\propto l^{-\alpha _{E,B}}$ with $\alpha
_{E,B}\simeq 1.4 - 1.5,$ in the portion of the Southern Galactic Plane
probed by the D97 Parkes survey. Although such slopes are close to the
values $\alpha _{E,B} ^{\mathrm {(dust)}}
\simeq 1.3 - 1.4$ found for thermal dust emission
(Prunet et al. 1998), later work cast serious doubts on the generality of
this behaviour, as far as synchrotron is concerned. For the APS of the
scalar $PI\equiv \sqrt{Q^2+U^2}$ somewhat steeper spectra, with $\alpha
_{PI} $ ranging from $1.7\pm 0.2$ to $1.9\pm 0.3$, are given for the
Galactic Plane by Baccigalupi et al. (2001a) and Giardino et al. (2001). The
former authors also find a much higher slope, $\alpha _{PI}=2.9\pm 0.2$ for $%
l\la  10^2,$ out of the Galactic Plane from the 1.4 GHz, low-resolution
survey of Brouw and Spoelstra (1976: BS76). This discrepancy would be
important when evaluating the synchrotron contamination at large scales;
it would thereby be rather unconfortable in view of 
the role of CMB polarization
harmonics with $l\approx 2-20$ in the separation of effects of the
primordial gravitational background from those of the secondary ionization
of the cosmic medium. [This is one of the objectives of the SPOrt
project, see e.g. Cortiglioni et al. (2001).] Note that the
situation is not quite clear for the total intensity $I$ either: Large
values $\alpha _I\simeq 3$ are supported by Bouchet and Gispert (1999) and
Giardino et al. (2001), but $\alpha _I\approx 2$ follows from analysis of
the Jodrell Bank 5-GHz interferometric survey by Giardino et al. (2000).

Here we extend our analysis of APS in the Galactic Plane considering the 2.7
GHz Effelsberg survey (Duncan et al. 1999: D99); further, we analyse several
patches at intermediate Galactic latitudes, $\left| b\right| \leq 20^{\circ
} $, from the 1.4 GHz Effelsberg survey (Uyaniker et al. 1999: U99), and
finally three regions from BS76, at 5 frequencies between 408 and 1411 MHz,
covering latitudes up to near the North Galactic Pole. Following Tucci et
al. (2001) and thereby extending the analysis of Baccigalupi et al. (2001a), 
we carefully distinguish between the APS providing a fuller
statistical description of the spin-2 polarization field (i.e., $C_{El}$ and 
$C_{Bl}$, from which $C_{Pl}=C_{El}+C_{Bl}$ is usually computed), and the $%
C_{PIl}$ spectrum which takes into account only the magnitude of the
polarization pseudovector. This 
is necessary because Tucci et al. (2001) find quite significant
differences between $C_{Pl}$ and $C_{PIl}$ both for a Gaussian polarization
field, as predicted in the standard scenario for CMB, and for synchrotron
radiation.

Our results confirm that slopes around $1.4 - 1.5$ in the range $100\leq
l\leq 800$ are preferred on the average, in both the Northern and 
the Southern
Galactic Plane for $C_{El}$, $C_{Bl}$ and $C_{Pl}$, although significant
fluctuations around theses values are found in $10^{\circ }\times 10^{\circ }
$ patches. The fluctuations however become stronger in small patches out of
the Galactic Plane, so that a preferred slope should not regarded as
meaningful for the U99 survey. 
In fact, the partial regularities that we
find for polarization APS do not support the usefulness of global (i.e.,
full sky) $C_{E,Bl}$ for a satisfactory description of the spatial
distribution of synchrotron: Local $C_{E,Bl}$ based on Fourier analysis are
much more suitable to this purpose. A 
consistent picture, however, emerges at all of the scales we investigated.
As a matter of facts, from the low-resolution BS76 survey we find moderate
slopes for $l\leq 70$, at all frequencies and even near the Galactic Pole.
Due to the limited range of $l$ available for the fits, we find strong
correlations between fitted parameters and, as a consequence, large error
bars; however there is no evidence for steep spectra, since the best values
lie in the range $\alpha _{E,B}\simeq 0.5 - 2.0$ almost everywhere and are
particularly small at frequencies where Faraday rotation is more important.
For the $C_{PIl}$ spectrum, the slope fluctuations are smaller
than for $C_{E,Bl}$. Our best
value for $\alpha _{PI}$ is $1.6 - 1.8$ at all resolutions and Galactic
latitudes, and all frequencies $\geq 1.4$ GHz. The difference between $%
C_{E,Bl}$ and $C_{PIl}$ is significant, although not large in the average.
The latter quantity should thereby be confronted with $C_{PIl}^{%
\mathrm{(CMB)}}$, not with the $C_{E,Bl}^{\mathrm{(CMB)}}$ spectra which are
popular among cosmologists, and this is properly done in the last Section
of this paper. 

All of the results that we provide for
polarization APS should not be affected by contamination 
from point sources;
regions with flat $C_{Il},$ on the other hand, are possibly dominated by
point sources in total intensity, and are used in this paper to derive upper
limits on their contribution to polarization APS, as discussed in
Section 3.

\section{Data analysis}

\subsection{Polarized synchrotron surveys}

Our first high-resolution analysis of Galactic-Plane synchrotron
polarization was performed on the 2.4 GHz Parkes survey (D97); this study,
using twelve $10^{\circ }\times 10^{\circ }$ square patches (Tucci et al.
2000), is extended here to the 2.7 GHz Effelsberg survey (1999: D99)
covering the additional Galactic longitude range $5^{\circ }\leq \ell \leq
74^{\circ }$ (see Table \ref{surveyspatches}) and providing six more $%
10^{\circ }\times 10^{\circ }$ patches. A belt extended about 54\% of the
Galactic Plane is thereby covered, although with two slightly different
frequencies. The FWHM resolutions are $10.4^{\prime }$ and $5.1^{\prime }$
respectively, sufficient to achieve angular scales up to $l\sim 10^3$. The
nominal rms noise in D97 is 8 mK for total power and 5.3 mK for polarization
(5.3 and 2.9 mK in some more sensitive areas), while in D99 the rms noise is
9 mK.

\begin{table}[!t]
\caption[]{Sky regions used for computation of APS}
\label{surveyspatches}
\begin{tabular}{lllll}
\hline
Ref. & $\nu $ (GHz) & $\mathrm {FWHM}$ 
& Galactic latitudes & Galactic longitudes \\ 
\hline
D97 & 2.4 & $10.4^{\prime }$ & $-5^{\circ }\leq b\leq 5^{\circ }$ & $%
-122^{\circ }\leq \ell \leq 5^{\circ }$ \\ \hline
D99 & 2.7 & $5.1^{\prime }$ & $-5^{\circ }\leq b\leq 5^{\circ }$ & $5^{\circ
}\leq \ell \leq 74^{\circ }$ \\ \hline
&  &  & $4^{\circ }\leq b\leq 20^{\circ }$ & $45^{\circ }\leq \ell \leq
55^{\circ }$ \\ 
&  &  & $3.5^{\circ }\leq b\leq 10^{\circ }$ & $140^{\circ }\leq \ell \leq
153^{\circ }$ \\ 
U99& 1.4 & $9.3^{\prime }$  & $3.8^{\circ }\leq b\leq 15^{\circ }$ 
& $190^{\circ }\leq \ell \leq
210^{\circ }$ \\
&  &  & $5^{\circ }\leq b\leq 15^{\circ }$ & $65^{\circ }\leq \ell \leq
95^{\circ }$ \\ 
&  & & $-15^{\circ }\leq b\leq -5^{\circ }$ & $%
70^{\circ }\leq \ell \leq 100^{\circ }$ \\ 
\hline
&  &  & $-10^{\circ }\leq b\leq 20^{\circ }$ & $120^{\circ }\leq \ell \leq
180^{\circ }$ \\ 
BS76 & $0.41-1.4$ & $2.3^{\circ }-0.6^{\circ }$ & $26.5^{\circ }\leq b\leq
62.5^{\circ }$ & $\ell _{\mathrm{c}}=60.5^{\circ },\Delta \psi =\pm
27^{\circ }$ \\ 
&  &  & $60.5^{\circ }\leq b\leq 84.5^{\circ }$ & $\ell _{\mathrm{c}%
}=30^{\circ },\Delta \psi =\pm 15^{\circ }$ \\ \hline
\end{tabular}
\end{table}

Our study is further extended to moderate Galactic latitudes analysing the
maps of the Effelsberg 1.4 GHz survey (U99), consisting of five regions
fairly close to the Galactic Plane, $-15^{\circ }\leq b\leq 20^{\circ }$.
The angular resolution is there $9.35^{\prime }$, and the rms noise is about
15 mK for total intensity and about 8 mK for linear polarization. The above
regions having different sizes, we extracted different numbers of 
patches from them, namely, 1, 2, 2, 3 and 3 patches
respectively, for regions corresponding to increasing row
numbers in the U99 sector of Table \ref
{surveyspatches}. The patch size is   
$10^{\circ }\times 10^{\circ }$ almost everywhere
(so that a portion of the first 
region is not used), but only 
$6^{\circ }\times 6^{\circ }$
in the second region. 

{\begin{figure}
\centerline{\epsfysize=6cm
\epsfbox{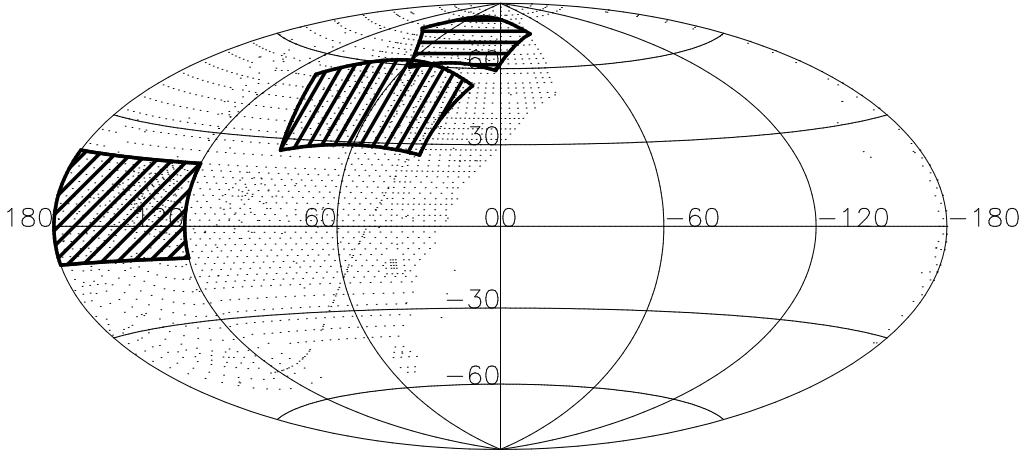}}
\caption{The selected patches in the BS76 survey. The patches
are the same at all frequencies; the dots give the locations
of the original BS76 measurements at 408 MHz.}
\label{threepatches}
\end{figure} }

{\begin{figure}
\centerline{\epsfysize=10cm
\epsfbox{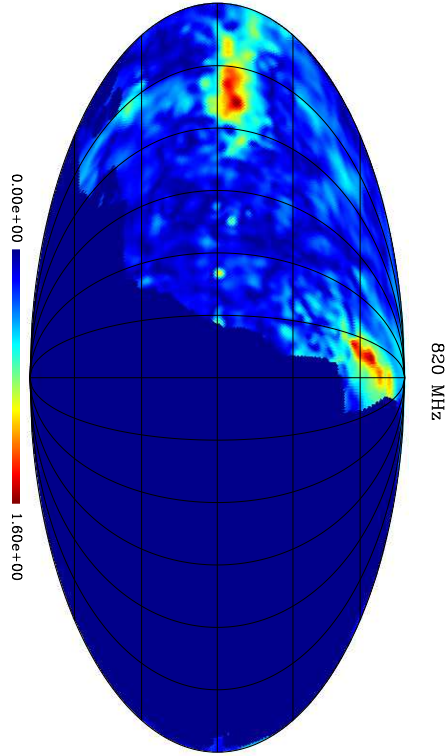}}
\caption{The total polarization ($PI$) field 
(in Kelvin degrees) for the BS76 survey
at 820 MHz, computed applying a Gaussian smoothing with 
$\mathrm {FWHM} = 3^{\circ }$
to the $Q$ \& $U$ fields.}
\label{sigbsp}
\end{figure} }

Finally we analysed the maps of Brouw and Spoelstra (1976: BS76) at five
frequencies between 408 and 1411 MHz. The corresponding beamwidths and noise
levels are summarized in Table~\ref{brouwspoe}. These maps cover a
substantial (more than 40\%) part of the sky, allowing us to compare results
at quite different latitudes up to near the Galactic Pole, although with
moderate resolutions. In order to get good sampling and large
signal-to-noise ratios we selected three rectangular regions, whose
locations are reported in the last three rows of Table~\ref{surveyspatches}
and in Fig.~\ref{threepatches}. As an example, we show the 820-MHz
total polarization signal in Fig.  \ref{sigbsp}.
The BS76 maps are affected by undersampling, and the grid spacing 
(typically larger
than the $\mathrm {FWHM}$) 
depends on the sky region and frequency. 
In the selected patches the average grid spacing is about $2^{\circ}$.
To make Fourier
analysis feasible we thereby constructed three new, evenly spaced grids
adopting a Gaussian smoothing function with dispersion $\sigma _{\mathrm{sm}%
}= \mathrm {FWHM} / \sqrt{8\ln 2}$ and $\mathrm {FWHM} = 3^{\circ }$. 
This procedure is straightforward near the
Galactic Plane, where geodesics orthogonal to meridians can be identified
with parallels with no appreciable errors. At high latitudes we constructed
the grids projecting orthogonal geodesics, evenly spaced with $\Delta
b=1.5^{\circ }$ at starting points, out of the central meridian of each
patch, having longitude $\ell =\ell _{\mathrm{c}}$ as reported in the last
two rows of Table~\ref{surveyspatches}. 
On such geodesics, extended respectively up to proper
lengths $\Delta \psi =\pm 27^{\circ }$ and $\pm 15^{\circ }$, we picked out
the centres of the Gaussian beams, with spacing $\Delta \psi ^{\prime
}=1.5^{\circ }$. (Since these centres do not lie on parallels, the latitude
ranges in the Table refer to central longitudes $\ell _{\mathrm{c}}$.) We
checked that for the chosen $\Delta \psi $ the proper distance between
corresponding points on \textit{neighbour} geodesics remains sufficiently
close to $\Delta \psi ^{\prime }$ even at the extremal longitudes $\ell
=\ell _{\mathrm{c}}\pm \Delta \psi $. This in fact provides an intrinsic
check for the applicability of flat-space concepts to celestial-sphere
patches.

\begin{table}
\caption[]{Parameters of the BS76 survey}
\label{brouwspoe}
\begin{tabular}{lll}
\hline
$\nu $ (MHz) & $\mathrm{FWHM}$ ($^{\circ }$) 
& Noise (K) \\ \hline
408 & 2.3 & 0.34 \\ 
465 & 2.0 & 0.33 \\ 
610 & 1.5 & 0.16 \\ 
820 & 1.0 & 0.11 \\ 
1411 & 0.6 & 0.06 \\ \hline
\end{tabular}
\end{table}

\subsection{Computation of angular spectra}

Because of the limited sky coverages, power spectra can be suitably obtained
from Fourier analysis instead of the standard spherical-harmonic approach
(Seljak 1997). The estimators for the power spectra of fields can be derived
by means of

\begin{equation}
C_{Xl}=\left[ \frac \Omega {N_l\mathbf{\,}}\sum_{\stackrel{\rightarrow }{l}}
X(\stackrel{\rightarrow }{l})X^{*}(\stackrel{\rightarrow }{l})\mathbf{-}%
w_X^{-1}\right] b_l^{-2}\mathbf{,}  \label{fouriercl}
\end{equation}
where $\Omega $ is the solid angle of the patch under analysis, $N_l$ the
number of Fourier modes in the interval around $l$ chosen for averaging, $%
w_X^{-1}$ the noise contribution and $b_l$ the beam function. Note that $X(%
\stackrel{\rightarrow }{l})$ is the discrete Fourier transform of the
smoothed field $X(\theta _i,\phi _i)$, a function of the bin coordinates
which can be any of the Stokes parameters $I$, $Q$ and $U$, but also
(considering the electric and magnetic parity polarization) $E$ and $B$. In
the small scale limit, electric and magnetic modes can in fact be given in
terms of $Q$, $U$ modes by means of a simple rotation in $l$--space:

\begin{eqnarray}
E(\stackrel{\rightarrow }{l}) &=&Q(\stackrel{\rightarrow }{l})\cos (2\phi _{%
\stackrel{\rightarrow }{l}})+U(\stackrel{\rightarrow }{l})\sin (2\phi _{%
\stackrel{\rightarrow }{l}}),  \nonumber \\
B(\stackrel{\rightarrow }{l}) &=&-Q(\stackrel{\rightarrow }{l})\sin (2\phi _{%
\stackrel{\rightarrow }{l}})+U(\stackrel{\rightarrow }{l})\cos (2\phi _{%
\stackrel{\rightarrow }{l}}).  \label{rotationeb}
\end{eqnarray}
Thus dealing with maps of the fields $Q$ and $U,$ we end up with $%
C_{El}$ and $C_{Bl}$, which are most
suitable for comparison with CMB spectra. Of course one could directly use $%
C_{Ql}$ and $C_{Ul}$, but these spectra are not rotationally invariant. The
APS describing the total power of the polarization field (taking into
account spatial variations of both magnitude and direction) is 
\begin{equation}
C_{Pl}=C_{Ql}+C_{Ul}=C_{El}+C_{Bl},\qquad l\geq 2.  \label{totalpower}
\end{equation}
Note the lower limit on $l$, arising from the spin-2 nature of linear
polarization (namely, of the $Q$, $U$ complex). 
Other authors have considered the APS
of the polarized intensity $PI=\sqrt{Q^2+U^2}$, which is defined for any $%
l\geq 0$. Equation (\ref{fouriercl}) still applies with $X=PI$.
The $PI$ field however 
does not provide a complete description of polarization, being 
only related to the                                         
magnitude of the polarization pseudovector. The equality $C_{PIl}
= C_{Pl}$  is warranted
only if the polarization angle 
is uniform inside all the survey area, and different spectral behaviours
arise even for Gaussian random fields (Tucci et al. 2001). 

In conclusion \emph{as many as 7 APS can be defined, without counting
cross-correlation and circular polarization
spectra}. Out of these, 4 are mutually independent, and
out of the latter, 3 are polarization APS.

The technique based on Eqs. (\ref{fouriercl}) and (\ref{rotationeb}), which
has been already applied by Tucci et al. (2000) on D97 data, is fairly
straightfoward for the small-size patches of high-resolution surveys. As
done in that work, it is implemented here with a cosine apodization to
suppress border effects, and with subtraction of mean values to suppress
aliasing. However, a careful analysis is required for BS76 maps. We have
already remarked in the previous Subsection that for high-latitude patches
we need geodesics to identify equally spaced grids to be used in Fourier
analysis; now we observe that we must also take reference frame effects
into account for $Q$ and $U$ fields. Such effects are especially important
for our last patch under analysis, whose border 
is very close to the Galactic Pole.

In particular, in order to smooth the $Q$ and $U$ fields we use the
following procedure: (i) at each point within a proper distance $\chi \leq
5\sigma _{\mathrm{sm}}$ from a Gaussian beam centre, we construct the
polarization vector $\stackrel{\rightarrow }{p}$, (ii) we perform a parallel
transport of $\stackrel{\rightarrow }{p}$ to the beam centre, obtaining 
\begin{equation}
p_\theta ^{\mathrm{(PT)}}=p_\theta \cos t+p_\phi \sin t,\quad p_\phi ^{%
\mathrm{(PT)}}=-p_\theta \sin t+p_\phi \cos t,  \label{partrans}
\end{equation}
with $t=\int \cos \theta \mathrm{d}\phi ,$ and (iii) from $p_{\theta ,\phi
}^{\mathrm{(PT)}}$we recover the parallel-transported Stokes parameters $Q^{%
\mathrm{(PT)}}$ and $U^{\mathrm{(PT)}}$ and apply the weight function $\exp
[-\chi ^2/(2\sigma _{\mathrm{sm}}^2)]$ to them, obtaining the
smoothed quantities
\begin{eqnarray}
Q_{\mathrm{sm}}^{\mathrm{(PT)}} &=&\frac{\sum_kQ_k^{\mathrm{(PT)}}\exp
[-\chi _k^2/(2\sigma _{\mathrm{sm}}^2)]}{\sum_k\exp [-\chi _k^2/(2\sigma _{%
\mathrm{sm}}^2)]},  \nonumber  
\\
U_{\mathrm{sm}}^{\mathrm{(PT)}} &=&\frac{\sum_kU_k^{\mathrm{(PT)}}\exp
[-\chi _k^2/(2\sigma _{\mathrm{sm}}^2)]}{\sum_k\exp [-\chi _k^2/(2\sigma _{%
\mathrm{sm}}^2)]}.  \label{qusmooth}
\end{eqnarray}
We checked that in practice the parallel transport could be made with
negligible variations on any reasonable path close to geodesics, and a very
convenient approximation is $t=\cos \theta _{\mathrm{ave}}\Delta \phi $,
with $\theta _{\mathrm{ave}}$ the average of polar angles at the starting
and end points. The smoothed fields $Q_{\mathrm{sm}}^{\mathrm{(PT)}}$ and $%
U_{\mathrm{sm}}^{\mathrm{(PT)}}$ are then used in Eq. (\ref{fouriercl}) for
the angular spectra of BS76 maps.

\section{Results}

\subsection{High-resolution surveys}

Angular spectra for total
intensity and E-parity polarization found in the
Galactic Plane are shown in Fig. \ref{duncan9799}. For most of the square
patches the curves are reasonably approximated by power laws, 
\begin{equation}
C_{Xl}=A_Xl^{-\alpha _X}.  \label{powerlaws}
\end{equation}
This holds not only for $X=I$ and $E$, but also for the $B$, $Q$, $U$, and $%
\,P$ spectra which, having similar behaviours, are not shown in the Figure. 
In Table \ref{tab:9799} we
report the results of spectral fits 
for $X=I$, $E$ and $B$, in the range $100\le
l\le 800$ for all of the 18 Galactic Plane patches.

{\begin{figure}
\centerline{\epsfysize=10cm
\epsfbox{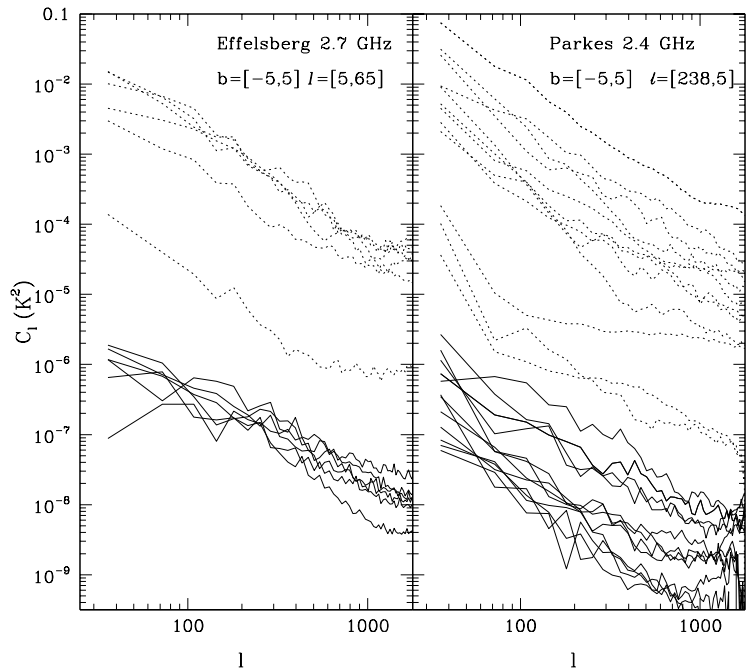}}
\caption{The APS for total intensity (dotted lines)
end $E$-mode polarization (solid lines), from 6
patches of the 2.7-GHz Effelsberg survey and 12 patches of the
2.4-GHz Parkes survey. }
\label{duncan9799}
\end{figure} }

\begin{table}[!t]
\caption[]{Best-fit parameters for APS from D97 and D99 surveys}
\label{tab:9799}
\begin{tabular}{llllll}
\hline
$\ell \,$($^{\circ }$)$^{\dagger}$ 
& $A_I$ (K$^2)$ & $\alpha _I$ & $A_E$ (K$^2)$ 
& $\alpha
_E $ & $\alpha _B$ \\ \hline
250 & 0.14$\times 10^{-3}$ & 0.96$\,_{-0.08}^{+0.12}$ & 0.13$\times 10^{-4}$
& 1.58$\,\pm $0.11 & 1.45$\,\pm $0.11 \\
260 & 0.20$\times 10^{-2}$ & 1.48$\,\pm $0.12 & 0.96$\times 10^{-5}$ & 1.48$%
\,\pm $0.12 & 1.32$\,\pm $0.10 \\ 
270 & 0.02 & 1.02$\,\pm $0.11 & 2.6$\times 10^{-3}$ & 1.79$\,\pm $0.12 & 1.25%
$\,\pm $0.12 \\ 
280 & 0.48$\times 10^{-4}$ & 0.44$\,_{-0.12}^{+0.08}$ & 0.44$\times 10^{-4}$
& 1.56$\,\pm $0.12 & 1.78$\,\pm $0.11 \\ 
290 & 1.8 & 1.43$\,\pm $0.11 & 0.32$\times 10^{-3}$ & 2.04$\,\pm $0.12 & 1.90%
$\,\pm $0.12 \\ 
300 & 0.07 & 1.22$\,_{-0.08}^{+0.12}$ & 0.15$\times 10^{-4}$ & 1.56$\,\pm $%
0.12 & 1.88$\,\pm $0.12 \\ 
310 & 5.1 & 1.89$\,\pm $0.12 & $0.28\times 10^{-5}$ & 1.11$\,\pm $0.11 & 0.96%
$\,_{-0.09}^{+0.12}$ \\ 
320 & 0.48 & 1.76$\,\pm $0.11 & 0.40$\times 10^{-4}$ & 1.49$\,\pm $0.11 & 
1.40$\,\pm $0.12 \\ 
330 & 0.21 & 1.46$\,\pm $0.09 & 0.22$\times 10^{-3}$ & 1.57$\,\pm $0.12 & 
1.74$\,_{-0.09}^{+0.15}$ \\ 
340 & 17.5 & 2.00$\,\pm $0.11 & 0.48$\times 10^{-4}$ & 1.31$\,\pm $0.11 & 
1.52$\,\pm $0.12 \\ 
350 & 3.5 & 1.64$\,\pm $0.11 & $0.92\times 10^{-6}$ & 0.85$%
\,_{-0.06}^{+0.13} $ & 1.12$\,\pm $0.12 \\ 
360 & 19.5 & 1.65$\,\pm $0.12 & 0.63$\times 10^{-4}$ & 1.28$\,\pm $0.11 & 
1.33$\,\pm $0.12 \\ 
\hline
10 & 3.40 & 1.60$\,\pm $0.12 & 0.21$\times 10^{-4}$ & 0.92$\,\pm $0.10 & 1.13%
$\,_{-0.12}^{+0.08}$ \\ 
20 & 124.4 & 2.18$\,\pm $0.07 & 0.39$\times 10^{-3}$ & 1.44$%
\,_{-0.07}^{+0.13}$ & 1.52$\,_{-0.15}^{+0.07}$ \\ 
30 & 11.5 & 1.79$\,\pm $0.10 & 0.88$\times 10^{-3}$ & 1.50$\,\pm $0.10 & 1.63%
$\,\pm $0.10 \\ 
40 & 0.10 & 1.21$\,\pm $0.12 & 0.25$\times 10^{-3}$ & 1.36$%
\,_{-0.06}^{+0.11} $ & 1.67$\,_{-0.10}^{+0.05}$ \\ 
50 & 0.31 & 1.28$\,_{-0.15}^{+0.21}$ & 0.80$\times 10^{-4}$ & 1.24$\,\pm $%
0.09 & 1.52$\,\pm $0.13 \\ 
60 & 0.65$\times 10^{-3}$ & 1.00$\,_{-0.15}^{+0.12}$ & 0.71$\times 10^{-3}$
& 1.68$\,\pm $0.07 & 1.82$\,\pm $0.10 \\ \hline
\end{tabular} 
\begin{minipage}[t]{5.in}
$^{\dagger}$Galactic longitude of the patch centre.
The patch belongs to D97 for $\ell=250^{\circ} - 360^{\circ}$, 
and to D99 for
$\ell=10^{\circ} - 60^{\circ}]$.
\end{minipage}
\end{table}

\begin{table}[!t]
\caption[]{Average parameters from Galactic Plane fits}
\label{avegp}
\begin{tabular}{llllll}
\hline
Survey \&\ method & $A_I$ (K$^2)$ & $\alpha _I$ & $A_E$ (K$^2)$ & $\alpha _E$
& $\alpha _B$ \\ \hline
D97, $\langle \alpha _X\rangle $ & $-$ & 1.37$\,\pm $0.44 & $-$ & 1.44$\,\pm 
$0.30 & 1.46$\,\pm $0.29 \\ 
D97, $\langle C_l\rangle $ & 2.2 & 1.60$\,\pm $0.13 & 0.12$\times 10^{-3}$ & 
1.53$\,\pm $0.11 & 1.43$\,\pm $0.12 \\ \hline
D99, $\langle \alpha _X\rangle $ & $-$ & 1.71$\,\pm $0.43 & $-$ & 1.40$\,\pm 
$0.23 & 1.57$\,\pm $0.19 \\ 
D99, $\langle C_l\rangle $ & 10.3 & 1.82$\,\pm $0.11 & 0.31$\times 10^{-3}$
& 1.39$\,\pm $0.11 & 1.55$\,\pm $0.12 \\ \hline
D97+D99, $\left\langle \alpha _X\right\rangle $ & $-$ & $1.48\pm 0.46$ & $-$
& $1.42\pm 0.30$ & $1.51\pm 0.26$ \\ 
D97+D99, $\left\langle C_\ell \right\rangle $ & 6.9 & $1.72\pm 0.10$ & 0.26$%
\times 10^{-3}$ & $1.40\pm 0.12$ & $1.54\pm 0.10$ \\ \hline
\end{tabular}
\end{table}

{\begin{figure}
\centerline{\epsfysize=13cm
\epsfbox{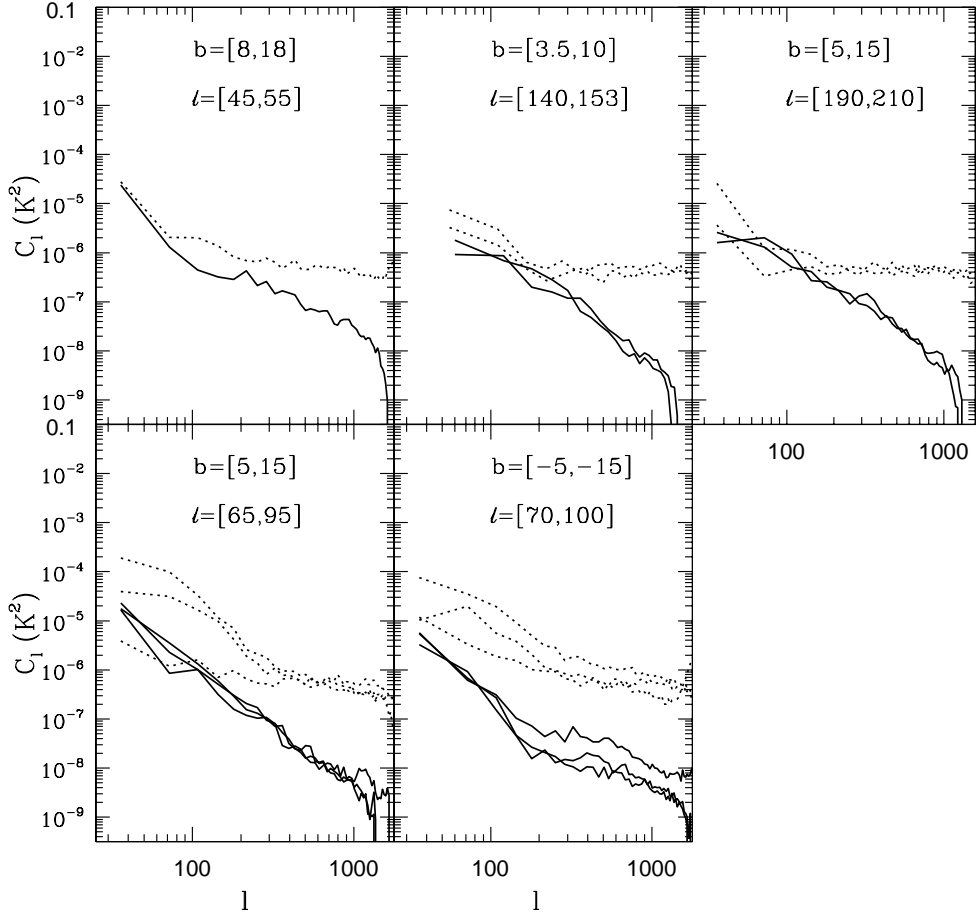}}
\caption{The APS for intensity (dotted lines)
end $E$-mode polarization (solid lines), from 5
regions of the 1.4-GHz Effelsberg survey. }
\label{uya99}
\end{figure}  }

\begin{table}[!t]
\caption[]{Best-fit parameters for APS from the U99 survey} 
\label{tab:u99}
\begin{tabular}{lllllll}
\hline
$\ell \,$($^{\circ }$) & $b\,$($^{\circ }$) & $A_I$ (K$^2)$ & $\alpha _I$ & $%
A_E$ (K$^2)$ & $\alpha _E$ & $\alpha _B$ \\ \hline
50 & 13 & 0.53$\times 10^{-5}$ & 0.37$\,_{-0.10}^{+0.13}$ & 0.16$\times
10^{-3}$ & 1.22$\,\pm $0.08 & 1.19$\,\pm $0.08 \\ \hline
143 & 7 & 0.35$\times 10^{-6}$ & 0.$\,^{+0.02}$ & 0.22 & 2.55$%
\,_{-0.17}^{+0.14}$ & 2.70$\,_{-0.25}^{+0.03}$ \\ 
150 & 7 & 0.50$\times 10^{-6}$ & 0.$\,^{+0.02}$ & 0.80$\times 10^{-1}$ & 2.38%
$\,_{-0.17}^{+0.06}$ & 2.10$\,_{-0.16}^{+0.23}$ \\ \hline
195 & 10 & 0.45$\times 10^{-6}$ & 0.$\,^{+0.02}$ & 0.35$\times 10^{-1}$ & 
2.28$\,\pm $0.08 & 2.32$\,\pm $0.08 \\ 
205 & 10 & 0.72$\times 10^{-6}$ & 0.$\,^{+0.11}$ & 0.70$\times 10^{-2}$ & 
1.99$\,_{-0.12}^{+0.08}$ & 1.98$\,\pm $0.10 \\ \hline
70 & 10 & 0.99$\times 10^{-4}$ & 0.82$\,\pm $0.15 & 0.73$\times 10^{-1}$ & 
2.41$\,\pm $0.11 & 2.39$\,\pm $0.12 \\ 
80 & 10 & 0.89$\times 10^{-3}$ & 1.13$\,\pm $0.13 & 0.90$\times 10^{-3}$ & 
1.71$\,\pm $0.15 & 1.79$\,\pm $0.19 \\ 
90 & 10 & 0.6$\times 10^{-6}$ & 0.$\,^{+0.13}$ & 0.11 & 2.48$%
\,_{-0.12}^{+0.02}$ & 2.23$\,\pm $0.13 \\ \hline
75 & -10 & 0.16$\times 10^{-5}$ & 0.16$\,\pm $0.13 & 0.23$\times 10^{-5}$ & 
0.87$\,\pm $0.08 & 1.50$\,\pm $0.12 \\ 
85 & -10 & 0.12$\times 10^{-4}$ & 0.87$\,\pm $0.12 & 0.36$\times 10^{-5}$ & 
1.00$\,\pm $0.11 & 1.17$\,\pm $0.12 \\ 
95 & -10 & 0.20$\times 10^{-4}$ & 0.81$\,\pm $0.11 & 0.47$\times 10^{-4}$ & 
1.20$\,\pm $0.09 & 0.63$\,_{-0.02}^{+0.07}$ \\ \hline
\end{tabular}
\end{table}

Total intensity spectra exhibit large variations in amplitude, by
more than three orders of magnitude. (The variations appear magnified
in the normalization parameter $A_I$, up to five orders of magnitude, 
because of a positive correlation with $\alpha _I$.)
The same feature is not found in
polarization spectra. This can be explained by observing that, while the
total power emission decreases fast when moving away from the Galactic
centre, the polarized component remains much more uniform changing the
Galactic longitude and latitude [see Figs. 4 and 6 in Duncan et al. (1995),
and Fig. 9 in D97]; D97 noticed a ``background component'' in the polarized
emission of about 20 mK, nearly constant over the entire survey. 
The distributions of the indices $\alpha
_X$, too, highlight some differences between total intensity and
polarization spectra: The values of $\alpha _I$ range between 0.4 and 2.2,
while the slopes of the polarization spectra in all patches remain
relatively close to the mean value $\simeq 1.4-1.5$. Moreover, low--emission
regions show very flat spectra in total intensity, while no meaningful
differences are found between high-- and low--polarized emission regions.

In Table \ref{avegp} we report the values of the average slopes (rows
labelled with $\langle \alpha _X\rangle $), as well as the fit parameters
from the average spectra (rows labelled by $\langle C_l\rangle $) for both
D97 and D99 surveys. The D97+D99 averages are computed rescaling the D99 data
to 2.4 GHz using a spectral index 
$\beta _{\mathrm {syn}}=-2.8$ (Platania et al. 1998). The agreement is very
satisfactory. On the other hand, the $C_{PIl}$ spectra turn out to be
somewhat steeper; the average slopes turn out to be $\alpha _{PI}\simeq
1.6-1.8$ (see Table \ref{aveslopes} in the next Subsection), and
fluctuations around these values are quite moderate. Although the difference
with the above results is not large, it has a statistical significance, and
is not surprising in the light of still large differences found in the arcmin
angular range by Tucci et al. (2001).

Fig. \ref{uya99} and Table \ref{tab:u99} report the results obtained from 5
intermediate--latitude regions of U99. (The E-mode spectra in the Figure again
are illustrative of the behaviour of the other polarization modes.) 
The intensity spectra are found to be
extremely flat ($\alpha _I<1$) and low, indicating that the diffuse emission
drops just out the Galactic plane. On the contrary, the polarization spectra
do not show a decrease in amplitude with respect to the other two surveys;
this means that the polarized background observed in D97 and D99 
should extend at
least up to $\ell \sim 15^{\circ }$. However, in U99 we observe large
differences in the slope of polarization spectra from region to region:
there are three patches with $\alpha _{E,\,B}>2$ and two with $\alpha
_{E,\,B}\sim 1$. Some of these regions cannot be considered as
typical; for example, the area $140^{\circ }\le \ell \le 153^{\circ
},\,4^{\circ }\le b\le 10.4^{\circ }$ lies within the so called ``fan
region'', and the two regions centered at $\ell =80^{\circ }$ show rather
complex structures. Such particularly large
differences may be due to Faraday rotation effects, which are
relevant at 1.4 GHz. However a significant dispersion in the results 
remains at higher frequencies, and undoubtedly shows how
local (rather than global) spectra will be important for foreground
subtraction in CMB polarization measurements.

It is interesting to observe that several U99 patches exhibit 
nearly flat intensity spectra, i.e.
$\alpha _I\simeq 0$, with a remarkably uniform  normalization. 
It can be argued that these patches are dominated (in total 
intensity) by extragalactic point
sources, which are expected indeed to 
have flat APS as
far as clustering can be neglected (Tegmark and Efstathiou 1996; Toffolatti
et al. 1998). We can thereby estimate
$C_{Il }^{\mathrm{(PS)}} \simeq  5\times
10^{-7}$ K$^2.$  To be quite conservative, we can take this number as an 
upper limit.  
From this limit, adopting a radio-source polarization
degree of 5\% [in agreement with De Zotti et al. (2000)] we get $C_{Xl }^{%
\mathrm{(PS)}}<1.3\times 10^{-9}$ K$^2$ for $X=P$, $PI$ at 1.4 GHz. Assuming
further a (frequency) spectral index 
$\beta _{\mathrm {RG}}=-2$ for radiogalaxies, we also
get $C_{Xl }^{\mathrm{(PS)}}<1.5\times 10^{-10}$ K$^2$ at 2.4 GHz. We
conclude that the contribution of point sources should be negligible in the
whole range $l \leq 800$ for all of our \textit{polarization} APS. 
This is consistent with our results: The reported $C_{E,Bl}$, 
being $\ga 5\times
10^{-9}$ K$^2$ for $l <1000 $ at 1.4 GHz, do not show
any flattening. Note
also that more stringent limits on $C_{Il }^{\mathrm{(PS)}}$ (and
therefore, on $C_{P,PIl }^{\mathrm{(PS)}}$) might be derived from
estimates of Tegmark et al. (1996) and Toffolatti et al. (1998). These
however would be less safe.

\subsection{Low-resolution survey}

{\begin{figure}[h!]
\centerline{\epsfysize=8cm
\epsfbox{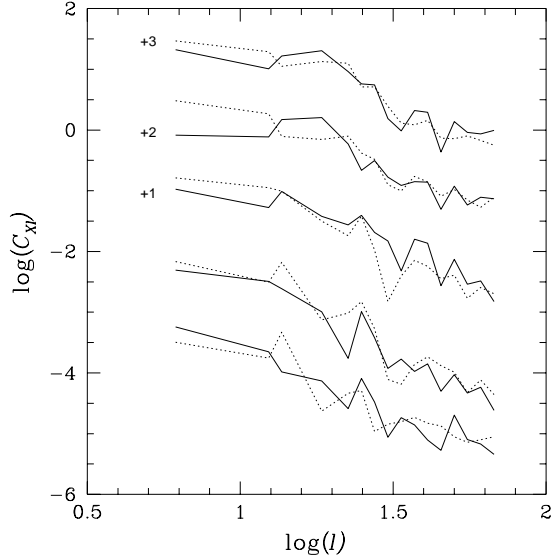}}
\caption{The APS for $E$-mode (solid lines) and $B$-mode (dotted lines)
for the BS76 patch centered at $b = 5^{\circ}$. 
From top to bottom,
the APS refer to frequencies of 408, 465, 610,
820 and 1411 MHz. Curves labelled by a number $n$ are shifted
by a factor $10^n$.}
\label{bs050}
\end{figure}  }

{\begin{figure}[h!]
\centerline{\epsfysize=8cm
\epsfbox{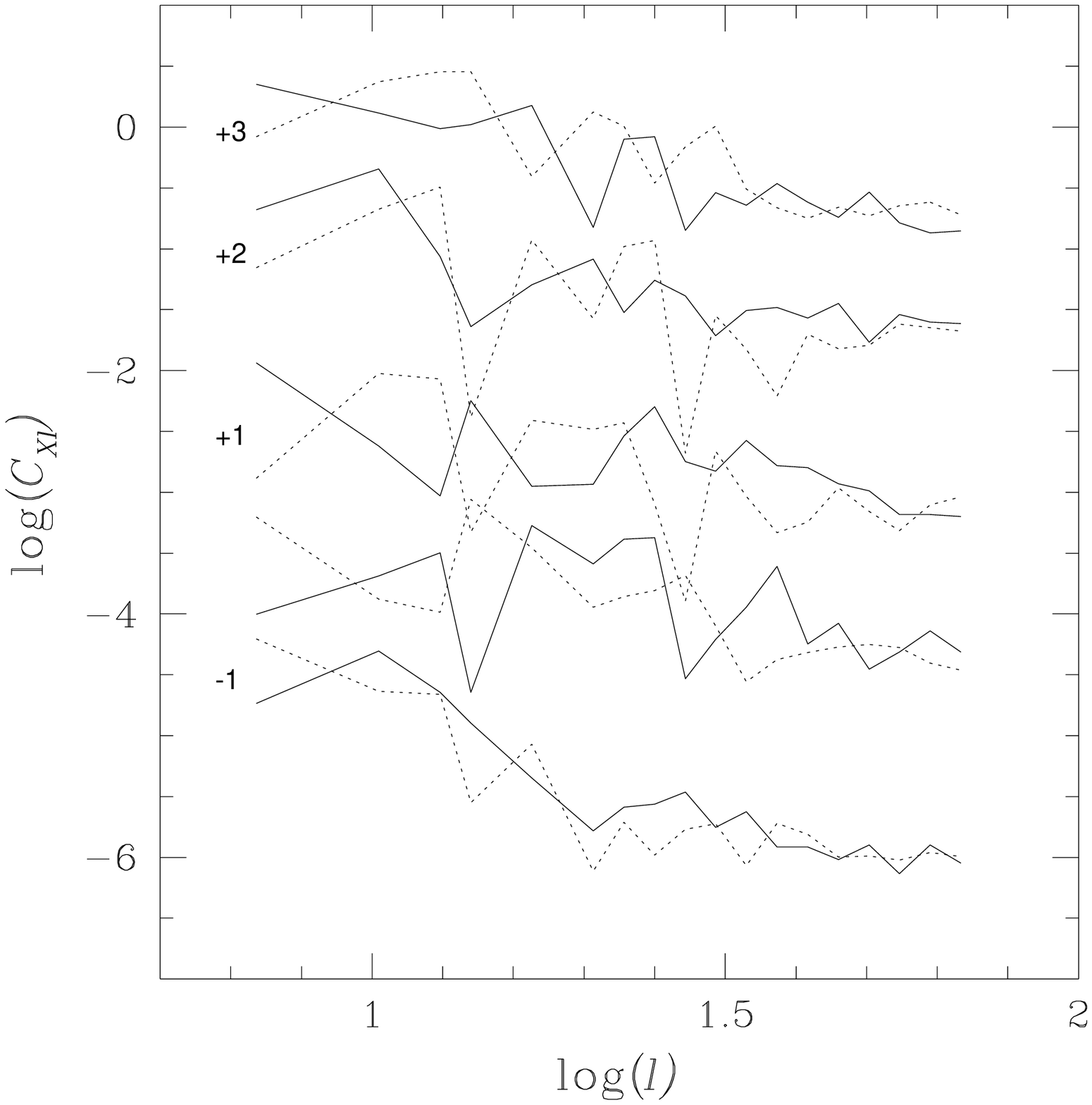}}
\caption{Same as Fig. \ref{bs050}
for the BS76 patch centered at $b = 44.5^{\circ}$. 
}
\label{bs445}
\end{figure}  }

{\begin{figure}[h!]
\centerline{\epsfysize=8cm
\epsfbox{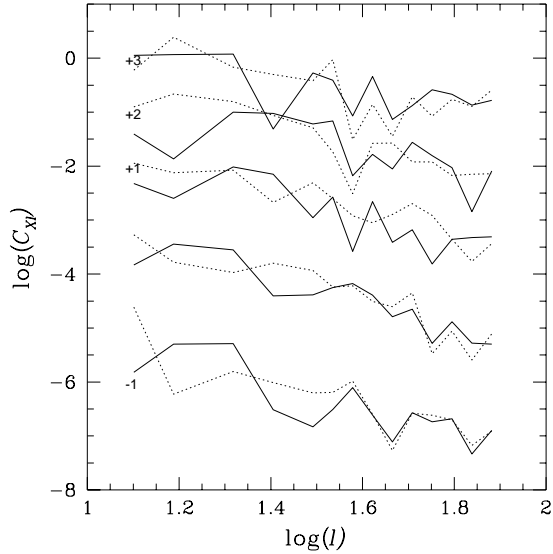}}
\caption{Same as Fig. \ref{bs050}
for the BS76 patch centered at $b = 72.5^{\circ}$. 
}
\label{bs725}
\end{figure}  }

For the BS76 survey we analysed the maps at all of the frequencies, but due
to the moderate resolution and the patch size, we could investigate only a
limited range of $l$. In particular, uncertainties in the beam angular
function for $l>\pi / \mathrm {FWHM}$ 
must exist in the original experiment; further,
adopting a Gaussian shape for our smearing function $\exp [-\chi ^2/(2\sigma
_{\mathrm{sm}}^2)]$ neglects the finite bin of the original measurements.
Therefore, it is advisable to limit ourselves to the range such that $%
b_l^{-2}\la  e$. In fact, the spectra computed by means of Eq. (\ref
{fouriercl}) turned out to increase for $b_l^{-2}\ga  e$, as can be
expected for the difficulty of an accurate calculation of noise for the
smoothed beams. 
Limiting ourselves to the range $ l \leq 70$, this effect
does not appear to visual inspection (see Figs. \ref{bs050}-\ref{bs725});
however, for the fits we adopt the modified function 
\begin{equation}
C_{Xl}=A_Xl^{-\alpha _X}+O_X b_l^{-2}.  \label{powermod}
\end{equation}
Here $O_X$, a parameter to be determined by the fit, is not intended to
describe the field $X$, but rather to account for the inaccurate \textit{a
priori} estimate of $w_X^{-1}$. The upper limits on point source APS given
in the previous Subsection, rescaled to the BS76 frequencies, exclude that
the $O_X b^{-2}$ term may mask important contributions from the above
sources. As a matter of facts, for 
$\beta _{\mathrm {RG}}=-2.3$ we find for instance $%
C_{X\ell }^{\mathrm{(PS)}}<4\times 10^{-7}$ K$^2$ for $X=P$, $PI$ at 408
MHz, some orders of magnitude below the measured APS.

{\begin{figure}[b!]
\centerline{\epsfysize=7cm
\epsfbox{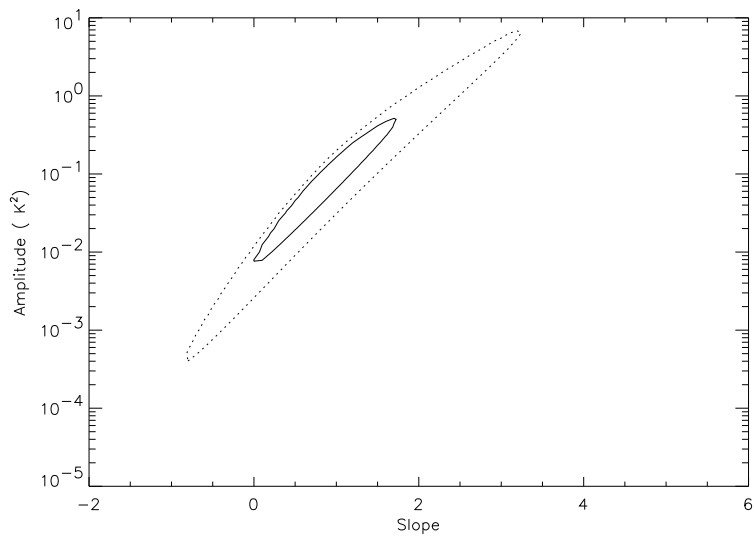}}
\caption{Confidence levels for $B$-mode parameters fitted at 408 MHz
for the BS76 patch centered
at $b = 5^{\circ}$. The
$\chi^2$ contours in the ($A_B$, $%
\alpha _B$) plane are reported at 1-$\sigma$ (solid lines) 
and 2-$\sigma$ levels (dotted), after minimization with respect to $O_B$.
}
\label{chicontour1408}
\end{figure} }

\begin{table}
\caption[]{Best-fit parameters for APS from the BS76 survey$^\dagger$}
\label{bsgppole}
\begin{tabular}{lllllllll}
\hline
$\nu $ (MHz) & $\mathrm{\log }A_E$ & $\alpha _E$ & $\mathrm{\log }A_B$ & $%
\alpha _B$ & $\mathrm{\log }A_P$ & $\alpha _P$ & $\mathrm{\log }A_{PI}$ & $%
\alpha _{PI}$ \\    \hline
408 & $-0.9_{-1.7}^{+0.4}$ & $1.0_{-1.4}^{+0.5}$ & $-0.7_{-1.3}^{+0.5}$ & $%
1.2_{-1.2}^{+0.5}$ & $-0.7_{-0.1}^{+0.1}$ & $0.8_{-0.2}^{+0.7}$ & $%
-0.2_{-0.3}^{+0.2}$ & $1.3_{-0.3}^{+0.1}$ \\ 
465 & $-2.1_{-0.9}^{+0.8}$ & $0.4_{-1.0}^{+0.3}$ & $-0.7_{-0.8}^{+0.4}$ & $%
1.5_{-0.9}^{+0.3}$ & $-0.9_{-1.7}^{+0.4}$ & $1.2_{-1.2}^{+0.5}$ & $%
-0.5_{-0.1}^{+0.3}$ & $1.1_{-0.2}^{+0.3}$ \\ 
610 & $-1.2_{-1.1}^{+0.4}$ & $1.2_{-1.0}^{+0.4}$ & $-1.0_{-0.5}^{+0.4}$ & $%
1.3_{-0.5}^{+0.6}$ & $-0.7_{-0.1}^{+0.2}$ & $1.1_{-0.2}^{+0.1}$ & $%
-0.5_{-0.1}^{+0.3}$ & $1.5_{-0.3}^{+0.2}$ \\ 
820 & $-1.3_{-0.5}^{+0.6}$ & $1.6_{-0.6}^{+0.7}$ & $-1.3_{-0.4}^{+0.3}$ & $%
1.4_{-0.3}^{+0.3}$ & $-0.9_{-0.1}^{+0.3}$ & $1.2_{-0.1}^{+0.6}$ & $%
-0.7_{-0.7}^{+0.6}$ & $1.5_{-0.5}^{+0.7}$ \\ 
1411 & $-2.0_{-1.2}^{+0.3}$ & $1.9_{-0.2}^{+0.4}$ & $-2.8_{-0.3}^{+0.3}$ & $%
1.2_{-0.3}^{+0.4}$ & $-1.6_{-0.7}^{+0.3}$ & $1.8_{-0.3}^{+0.5}$ & $%
-2.0_{-0.7}^{+0.5}$ & $1.8_{-0.9}^{+0.7}$ \\ \hline

408 & $-1.8_{-0.7}^{+0.3}$ & $1.2_{-0.7}^{+0.4}$ & $-1.5_{-0.8}^{+0.4}$ & $%
0.6_{-0.4}^{+0.6}$ & $-0.8_{-1.0}^{+1.1}$ & $1.1_{-0.8}^{+1.0}$ & $%
-1.1_{-0.5}^{+0.9}$ & $1.3_{-1.0}^{+1.1}$ \\ 
465 & $-1.2_{-0.5}^{+0.6}$ & $1.5_{-0.8}^{+0.6}$ & $-1.7_{-0.9}^{+0.6}$ & $%
1.3_{-0.9}^{+0.8}$ & $-1.0_{-0.8}^{+0.2}$ & $1.7_{-1.1}^{+0.3}$ & $%
-1.0_{-0.4}^{+1.0}$ & $1.3_{-0.3}^{+1.2}$ \\ 
610 & $-1.7_{-1.2}^{+1.4}$ & $1.3_{-0.6}^{+1.5}$ & $-2.5_{-1.3}^{+0.5}$ & $%
1.1_{-0.9}^{+0.4}$ & $-1.9_{-1.2}^{+0.2}$ & $1.1_{-0.9}^{+0.4}$ & $%
-1.2_{-0.6}^{+1.6}$ & $1.6_{-0.9}^{+1.0}$ \\ 
820 & $-2.3_{-2.1}^{+0.5}$ & $1.5_{-1.8}^{+1.0}$ & $-2.0_{-1.8}^{+0.8}$ & $%
1.2_{-1.8}^{+1.0}$ & $-1.8_{-1.5}^{+0.7}$ & $1.1_{-0.8}^{+0.9}$ & $%
-1.9_{-1.3}^{+1.0}$ & $1.7_{-0.6}^{+1.0}$ \\ 
1411 & $-2.5_{-0.8}^{+0.3}$ & $1.4_{-0.7}^{+0.5}$ & $-1.4_{-0.6}^{+0.7}$ & $%
2.2_{-0.8}^{+1.0}$ & $-1.5_{-0.6}^{+0.9}$ & $1.9_{-0.7}^{+1.0}$ & $%
-2.0_{-0.7}^{+0.9}$ & $1.7_{-0.7}^{+0.9}$ \\ \hline

408 & $-2.0_{-1.6}^{+0.7}$ & $0.8_{-1.3}^{+0.4}$ & $-2.2_{-1.2}^{+0.5}$ & $%
0.5_{-0.9}^{+0.4}$ & $-1.4_{-0.6}^{+0.4}$ & $0.4_{-0.3}^{+0.3}$ & $%
-0.4_{-0.7}^{+1.6}$ & $1.1_{-0.3}^{+1.2}$ \\ 
465 & $-1.7_{-1.0}^{+1.4}$ & $0.6_{-0.5}^{+1.4}$ & $-2.0_{-0.2}^{+0.1}$ & $%
0.5_{-0.1}^{+0.1}$ & $-1.7_{-1.7}^{+0.3}$ & $0.3_{-1.1}^{+0.3}$ & $%
-0.3_{-0.7}^{+1.0}$ & $0.9_{-0.5}^{+0.7}$ \\ 
610 & $-3.7_{-1.0}^{+1.2}$ & $-0.4_{-0.8}^{+1.2}$ & $-1.8_{-0.9}^{+3.5}$ & $%
1.0_{-0.8}^{+3.0}$ & $-1.4_{-1.8}^{+1.4}$ & $1.2_{-1.9}^{+1.3}$ & $%
-1.5_{-0.1}^{+2.4}$ & $1.2_{-0.9}^{+2.2}$ \\ 
820 & $-2.1_{-1.8}^{+0.8}$ & $1.2_{-2.0}^{+0.5}$ & $-1.7_{-2.1}^{+1.1}$ & $%
1.5_{-1.5}^{+0.9}$ & $-1.4_{-0.8}^{+0.7}$ & $1.0_{-0.6}^{+0.5}$ & $%
-1.5_{-0.3}^{+1.2}$ & $1.2_{-0.4}^{+0.8}$ \\ 
1411 & $-3.0_{-2.0}^{+0.5}$ & $1.4_{-2.1}^{+0.6}$ & $-1.3_{-1.7}^{+1.0}$ & $%
2.1_{-1.4}^{+1.0}$ & $-1.5_{-1.5}^{+1.4}$ & $1.9_{-1.3}^{+1.1}$ & $%
-2.0_{-1.8}^{+0.5}$ & $1.8_{-1.5}^{+1.0}$ \\ \hline
\end{tabular} 
\begin{minipage}[t]{6.in}
$^{\dagger}$Three sets of 5 frequencies refer to 3 patches, with increasing
Galactic latitudes (cfr. Table \ref{surveyspatches}).
\end{minipage}
\end{table}

\begin{table}[!t]
\caption[]{Average slopes of  $C_{Pl}$ and $C_{PIl}$ spectra}
\label{aveslopes}
\begin{tabular}{llll}
\hline
$\nu $ (MHz) & $l$-range & $\alpha _P$ & $\alpha _{PI}$ \\ \hline
408 & $\leq 70$ & $0.56\pm 0.24$ & $1.29\pm 0.19$ \\ 
465 & $\leq 70$ & $1.05\pm 0.43$ & $1.09\pm 0.22$ \\ 
610 & $\leq 70$ & $1.10\pm 0.14$ & $1.50\pm 0.24$ \\ 
820 & $\leq 70$ & $1.14\pm 0.28$ & $1.43\pm 0.37$ \\ 
1411 & $\leq 70$ & $1.82\pm 0.34$ & $1.76\pm 0.51$ \\ 
2417 & $100 - 800$ & $1.48 \pm 0.12 $ & $1.68\pm 0.30$ \\
2695 & $100 - 800$ & $1.50 \pm 0.11 $ & $ 1.59 \pm 0.12$ \\  \hline
\end{tabular}
\end{table}

The results of the fits are given in Table \ref{bsgppole}. The
declared uncertainties, which are 1-$\sigma $ 
errors on individual parameters
computed from $\chi ^2$ fields, are large because of correlations in
3-parameter fits. In particular, we find strong, positive correlations
between $A_X$ and $\alpha _X$, as illustrated by an example in Fig. \ref
{chicontour1408}, which refers to the 408-MHz low-latitude patch. 
This Figure shows
$ \chi ^2$ iso-contours  that we found in the ($A_X$, $%
\alpha _X$) plane after minimization with respect to $O_X$. In spite of 
the above
uncertainties, we find moderate slopes for $\alpha _E$ and $\alpha _B$,
generally in the range $1 - 2$. At low frequencies, $\nu \leq 610$ MHz
where Faraday rotation is larger, we occasionally find some slopes $<1$, and
in one case (near the Galactic Pole in the 610 MHz map) a best value $\alpha
_E<0.$ We should remark that due to the finite resolution of our sampling in
($A_X$, $\alpha _X$, $O_X$) space, and the existence of narrow, elongated
valleys with $ \chi ^2$ near minimum, the best values reported in Table 
\ref{bsgppole} are less significant than the full
extension of the confidence regions, described by the error bars in the
Tables. However, none of the above considerations would change 
if the centres of
such confidence regions were considered instead of the quoted minimum points.
Thus there is sufficient evidence that polarization APS are somewhat flatter
at low frequencies.

The angular spectral behaviour does not exhibit any clear dependence on
Galactic latitude at these moderate resolutions. It makes sense therefore to
average over our three patches from the BS76 survey. Rows 1 to 5 of Table 
\ref{aveslopes} gives the average slopes $\alpha _P$ and $\alpha _{PI}$ for
each BS76 frequency. At low frequencies, where Faraday rotation becomes more
and more important, we observe a gradual flattening of both spectra; the
deficit of steepness is more apparent for $\alpha _P$, in agreement with the
considerations of Tucci et al. (2001). On the other hand, at 1411 GHz where
Faraday rotation is less important (but not negligible at all), there is no
evidence for any difference between $\alpha _P$ and $\alpha _{PI}$ due to
the large error bars. Also, the result quoted for $\alpha _{PI}$ at 1411 MHz
is quite consistent to those found with higher resolution in the Galactic
Plane at 2.4 and 2.7 GHz (cfr. the last two rows in the Table).

\section{Conclusion}

{\begin{figure}[h!]
\centerline{\epsfysize=8cm
\epsfbox{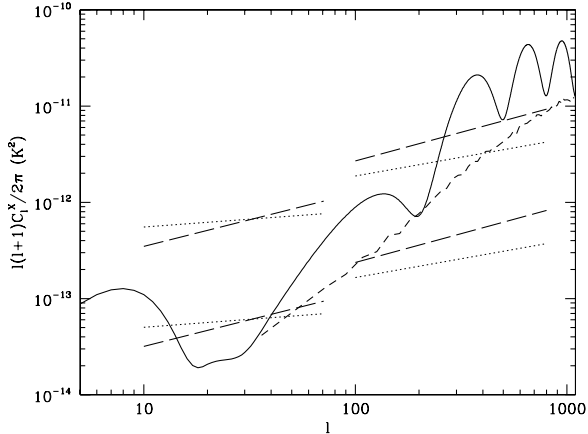}}
\caption{CMB and synchrotron APS. The CMB $C_{El}$ and  $C_{PIl}$
angular spectra, computed for a sCDM model with a reionization
optical depth $\tau _{\mathrm{ri}} = 0.1$, are represented 
by the solid and short-dashed curves
respectively. The corresponding synchrotron spectra are 
respectively given
by the long-dashed and dotted lines; these are extrapolated
to 60 (upper lines) and 90 GHz (lower), assuming 
a frequency spectral
index $\beta _{\mathrm {syn}} = - 3$. See the text for details.  
}
\label{cmbsyn}
\end{figure}  }

The regularities found for polarization APS in this work are not sufficient
to establish the usefulness of global (i.e., full sky) $C_{Xl}$ for
a satisfactory description of the spatial distribution of synchrotron
itself, and {\it a fortiori} for the
separation of Galactic synchrotron from CMB at higher frequencies. 
On the contrary, the local $C_{Xl}$ based on Fourier analysis are
much more suitable for a fuller description of angular structure. This
makes all-sky surveys of polarized synchrotron quite necessary.

Quite significant fluctuations are found for parameters fitted
in $10^{\circ }\times 10^{\circ }$ patches. When we average over larger
regions, the most stable behaviour is found
for the polarized intensity spectrum $C_{PIl}$
which, for all of the surveys analysed by us at $\nu \geq 1.4$ GHz,
everywhere exhibits a slope $\alpha _{PI}=1.6 - 1.8.$ This holds
in the full range $10\la  l\leq 800$, although we have large
error bars for the range $l\leq 70$ investigated on BS76 maps. In spite of
the more intricate situation for the other APS $C_{Xl}$ ($X=E$, $B$ and $P$%
), we can state the following:

\begin{itemize}
\item  In the Galactic Plane, the slopes of electric and magnetic parity APS
are quite moderate at 2.4 and 2.7 GHz, with averages $\alpha _E\simeq \alpha
_B\simeq 1.4 - 1.5$ in the range $100\leq l\leq 800$.

\item  Local fluctuations do not allow us to establish equally significant
average slopes out of the Galactic Plane in the same angular range.

\item  At lower resolution, $l\leq 70$, large correlation between fit
parameters cause large error bars; however, the best-fit slopes $\alpha _E$
and $\alpha _B$ stay in the range $0.5 - 2$ for almost all frequencies (in
the range $0.4 - 1.4$ GHz) and Galactic latitudes, and are quite
inconsistent with values around 3.
\end{itemize}

The above behaviours of polarization APS should be attributed to Galactic
synchrotron, with no appreciable contamination from point sources. On the
other hand, the total intensity APS may be locally dominated by sources,
when they exhibit small amplitude and slope $\alpha _I$ close to zero.

Our results resolve the seeming discrepancy with other authors for the
angular range $100\leq l\leq 800$, showing that investigators simply have to
carefully consider which polarization APS is actually being computed.
Our Galactic
Plane result, $\alpha _E\simeq \alpha _B\simeq 1.4 - 1.5,$ is very close to
results found for the polarization APS of thermal dust (Prunet et al.
1998); this number is maybe deeply
connected to Galactic structure.
On the other hand, we do not confirm the high spectral slope 
found by Baccigalupi et al. (2001a) for $%
C_{PIl}$ at smaller $l$ on three BS76 patches 
at 1.4 GHz. Our BS76 patches however
are different from theirs, being expressely chosen to have a larger
signal-to-noise ratio. Our best value $\alpha _{PI} = 1.76 \pm
0.51$, which arises from averaging over 1.4-GHz BS76 patches, 
is consistent with our results at all resolutions and frequencies
$\geq 1.4$  GHz. The latter also agree  with results obtained by
Tucci et al. (2001) in the arcminute range. The slope of starligth  
APS, $\alpha _{PI} ^{\mathrm {(star)}} \simeq 1.5$ (Fosalba et al. 2001), 
is closer to our $\alpha _{E,B} $.

Generally speaking, the $PI$ and $E$ \&\ $B$ fields contain 
different physical information. Since
$PI$ does not carry any information on the polarization angle,
its spectrum cannot properly describe some related effects 
like, for example, beam depolarization; it should be used
carefully, keeping in mind that it does not
provide a complete description 
of the polarization field. 

In particular, if $C_{PIl}$  is 
extrapolated
to the cosmological window, it is important to 
make a proper comparison with the 
corresponding APS of CMB. Figure \ref{cmbsyn} compares both
$C_{El}$ and $C_{PIl}$ spectra of synchrotron and CMB.
The reported CMB E-mode spectrum is the output of CMBFAST 
for a sCDM model
with a reionization
optical depth $\tau _{\mathrm{ri}} = 0.1$. 
The corresponding $C_{PIl}^{\mathrm {(CMB)}}$ 
spectrum is obtained through
simulations in a $10^{\circ} \times 10^{\circ}$ box,
with the mean value being
subtracted off. [Simulations
of $C_{El}^{\mathrm {(CMB)}}$ 
in the same box reproduced the CMBFAST output to
a satisfactory extent, see Tucci et al. (2000) for details.] 
The synchrotron APS in the Figure     
are extrapolated to 60 and 90 GHz  
from the Galactic Plane average spectra (6th row  
in Table \ref{avegp})
for $l \geq 100$, and from BS76 patch centered at $b
= 44.5^{\circ }$ for $l \leq 70$.  In both cases, 
the computed normalization applies to high-emission regions and
is not expected to be representative of the whole sky.
For the extrapolations to high frequencies we assume a spectral
index $\beta _{\mathrm {syn }} = - 3$ (Platania et al., 1998). 
The results taken altogether offer a
quite consistent picture. They imply that
in high emission regions, whathever 
polarization APS is chosen, the synchroton polarized
foreground should be comparable
to CMB at 60 GHz even at large $l$.
At 90 GHz the expected scenario looks more favourable
for CMB experiments, both at small and large angular scales. 
Reionization effects on CMB should  
be investigated at $l \sim  10$ by means of $C_{El}^{\mathrm {(CMB)}}$
rather  than $C_{PIl}^{\mathrm {(CMB)}}$.
From the inspection of Fig. \ref{cmbsyn}, and recalling that
the height of the low-$l$ peak of  $C_{El}^{\mathrm {(CMB)}}$
is approximately proportional $\tau _{\mathrm{ri}}^2$,  
we infer that the 90-GHz CMB signal from reionization should prevail
on synchrotron at least for $\tau  _{\mathrm {ri}}\ga 0.05$.
We finally remark that
in the $PI$ spectrum the
``DC'' signal $C_{PI0}^{\mathrm {(CMB)}}$ might be interesting, but
Fourier analysis is not appropriate to this purpose.  

\section*{Acknowledgments}

We thank T.A. Spoelstra for kindly providing
the BS76 data. This work was performed within the SPOrt collaboration,
and was financially supported by the Italian Space Agency (ASI).

\end{document}